\documentclass[aps,prl,twocolumn,amsmath,amssymb,nofootinbib,showpacs,superscriptaddress]{revtex4}
\usepackage[english]{babel}
\usepackage{latexsym}
\usepackage{graphics}
\usepackage{epsfig}
\usepackage{color}
\usepackage{bm}
\usepackage{amsmath}
\usepackage{amssymb}

\begin{document}

\title{Atom-dimer scattering and long-lived trimers in fermionic mixtures}

\author{J.~Levinsen}
\affiliation{\mbox{Laboratoire Physique Th\'eorique et Mod\`eles Statistique, Universit\'e Paris Sud, CNRS, 91405~Orsay, France}}
\author{T.~G.~Tiecke}
\affiliation{\mbox {Van der Waals-Zeeman Institute,University of Amsterdam,
Valckenierstraat 65/67, 1018 XE Amsterdam, The Netherlands}}
\author{J.~T.~M.~Walraven}
\affiliation{\mbox {Van der Waals-Zeeman Institute,University of Amsterdam,
Valckenierstraat 65/67, 1018 XE Amsterdam, The Netherlands}}
\author{D.~S.~Petrov}
\affiliation{\mbox{Laboratoire Physique Th\'eorique et Mod\`eles Statistique, Universit\'e Paris Sud, CNRS, 91405~Orsay, France}}
\affiliation{Russian Research Center Kurchatov Institute, Kurchatov Square, 123182 Moscow, Russia}

\date{\today}

\begin{abstract}
We consider a heteronuclear fermionic mixture on the molecular side of an interspecies Feshbach resonance and discuss atom-dimer scattering properties in uniform space and in the presence of an external confining potential, restricting the system to a quasi-2D geometry. We find that there is a peculiar atom-dimer $p$-wave resonance which can be tuned by changing the frequency of the confinement. Our results have implications for the ongoing experiments on Lithium-Potassium mixtures, where this mechanism allows for switching the $p$-wave interaction between a K atom and Li-K dimer from attractive to repulsive, and forming a weakly bound trimer with unit angular momentum. We show that such trimers are long-lived and the atom-dimer resonance does not enhance inelastic relaxation in the mixture, making it an outstanding candidate for studies of $p$-wave resonance effects in a many-body system.

\end{abstract}

\maketitle

The last several years have seen enormous advances in the field of ultracold fermionic gases. The main progress has been made in studies of the crossover from the Bardeen-Cooper-Schrieffer superfluidity to Bose-Einstein condensation of molecules (BCS-BEC crossover) in mixtures of two hyperfine states of the same atom (either Li or K) close to an intercomponent Feshbach resonance \cite{Review}. Tunability of these systems has allowed one to investigate the effect of a population imbalance on the BCS-BEC crossover \cite{Zwierlein2006,Partridge2006}, study vortices in strongly interacting rotating fermionic superfluids \cite{Zwierlein2005}, and explore other exotic phenomena. The BCS-BEC crossover in a mixture of different fermionic species and the formation of weakly bound heteronuclear bosonic dimers is now the subject of very active investigation. Naturally, most of the experimental progress has been made for the alkali mixture of $^6$Li and $^{40}$K. In particular, Taglieber and co-workers have cooled the system to quantum degeneracy \cite{Taglieber2007}, interspecies Feshbach resonances have been studied in a non-degenerate system \cite{Wille2008,Tiecke2009}, and, recently, the formation of weakly bound Li-K dimers at two of these resonances has been reported \cite{Voigt2009,Schreck2009}. The association of atoms into weakly bound dimers is an important step toward obtaining ultracold dipolar molecules, as has recently been demonstrated for fermionic $^{87}$Rb-$^{40}$K molecules at JILA \cite{Jin2008}. The same technique can be used to transfer bosonic $^6$Li-$^{40}$K dimers to the ground rovibrational state with a large dipole moment \cite{Aymar2005}. 

Although the dipole moment of weakly bound heteronuclear dimers is negligible, they can strongly interact by exchanging light atoms. For sufficiently large mass ratios this exchange interaction dominates over the kinetic energy of the dimers and can even lead to their crystallization \cite{Petrov2007}. On the few-body level the role of the mass imbalance is distinctly visible considering the problem of two identical heavy fermions of mass $M$ interacting resonantly with a light atom of mass $m$. For mass ratios $M/m>13.6$ an attractive effective potential mediated by the exchange of the light atom dominates over the centrifugal barrier for the identical heavy fermions. This leads to the Efimov effect - the existence of an infinite number of bound trimer states \cite{Efimov1973}. For smaller mass ratios the centrifugal barrier is dominant. It excludes the Efimov effect and also suppresses recombination processes requiring three atoms to approach each other to very short distances, which is crucial for the collisional stability of the gas \cite{Lifetime}. Amazingly, $M/m=13.6$ is not the only critical mass ratio for this system. Recently, Kartavtsev and Malykh \cite{Kartavtsev2007} have shown that for positive interspecies scattering length $a$ and for $M/m>8.2$ there is a weakly bound {\it not} Efimovian trimer state, which for smaller mass ratios turns into a $p$-wave atom-dimer scattering resonance. 

In this Letter we consider a heteronuclear fermionic mixture with $M/m<8.2$ and study the scattering of a heavy atom by a weakly bound heteronuclear molecule, pointing out not only the practical relevance but also uniqueness of the $^6$Li-$^{40}$K case ($M/m\approx 6.64$). We calculate $s$- and $p$-wave atom-dimer scattering phase shifts as functions of the collision energy taking into account the finite width of the interspecies Feshbach resonance. We show that for sufficiently small detunings the $p$-wave atom-dimer interaction is resonant and should dominate the dynamics of the atom-dimer mixture in the ultra-cold regime. We also show that the atom-dimer $p$-wave resonance can be tuned by confining the system to a quasi-2D geometry, and one can even turn it into a real trimer state. Finally, we predict that the resonant $p$-wave atom-dimer interaction does not amplify the collisional relaxation in the system. In particular, the lifetime of a trimer can exceed several seconds. The relaxation in $s$-wave atom-dimer and dimer-dimer collisions, more dangerous in the case of a narrow Feshbach resonance, can also be suppressed by decreasing the magnetic field detuning.

In order to solve the three-atom problem at hand we use the zero-range model, which is well justified for ultracold gases where the van der Waals range of the interatomic potential, $R_e$, is orders of magnitude smaller than all other lengthscales. The method has been applied to three unequal-mass fermions in free space near a wide Feshbach resonance \cite{Petrov2003}. The effect of the finite width of the resonance is described in Ref.~\cite{Petrov_v2_2004} in the case of three identical bosons. Mora and co-workers \cite{Mora2004} have included an external quasi-1D confinement in the treatment of the three equal-mass fermion problem. The synthesis of these techniques is outlined below.

Consider a pair of identical heavy fermions of mass $M$ interacting with a light atom of mass $m$ via an $s$-wave Feshbach resonance. We assume that the atoms are harmonically confined in the $z$-direction with the single particle Hamiltonians $\hat H_i= -\nabla_{{\bf R}_i}^2/2 M+M\Omega^2 Z_i^2/2$ and $\hat h= -\nabla_{\bf r}^2/2 m+m\omega^2 z^2/2$, where we set $\hbar=1$. In the zero-range approximation the interaction acts only on the ``boundaries'' ${\bf r = R_1}$ and ${\bf r = R}_2$ leading to the $1/|{\bf r - R}_i|$-singularities in the 3-body wavefunction $\Psi({\bf R}_1,{\bf R}_2,{\bf r})$. We introduce an auxiliary function $f({\bf R}_1,{\bf R}_2)=\lim_{{\bf r \rightarrow R}_1}4\pi |{\bf r - R}_1| \Psi({\bf R}_1,{\bf R}_2,{\bf r})$, which, in fact, contains all the information about the state of the system as $\Psi$ can be retrieved by solving the equation
\begin{eqnarray}\label{Schr}
(\hat H_1+\hat H_2+\hat h-E)\Psi & = &f({\bf R}_1,{\bf R}_2)\delta ({\bf r - R}_1)/2\mu \nonumber \\
&&\hspace{-4mm}- f({\bf R}_2,{\bf R}_1)\delta ({\bf r - R}_2)/2\mu,
\end{eqnarray} 
where $\mu=mM/(m+M)$ is the reduced mass. Below the three-atom threshold, $E<\Omega+\omega/2$, the mapping $f\rightarrow \Psi$ is unambiguous and is given by an integral operator, the kernel of which is obtained from the Green function of the left hand side of Eq.~(\ref{Schr}). In the limit ${\bf r} \rightarrow {\bf R}_1$ this integral expression for $\Psi$ reduces to the form
\begin{equation}\label{PsiLimit}
\Psi({\bf R}_1,{\bf R}_2,{\bf r}) \rightarrow (1/|{\bf r - R}_1|-\hat K) f({\bf R}_1,{\bf R}_2)/4\pi,
\end{equation} 
where the integral operator $\hat K$ gives a regular contribution in the considered limit. 

\begin{figure}[b]
\vspace{-2mm}
\includegraphics[width=\hsize]{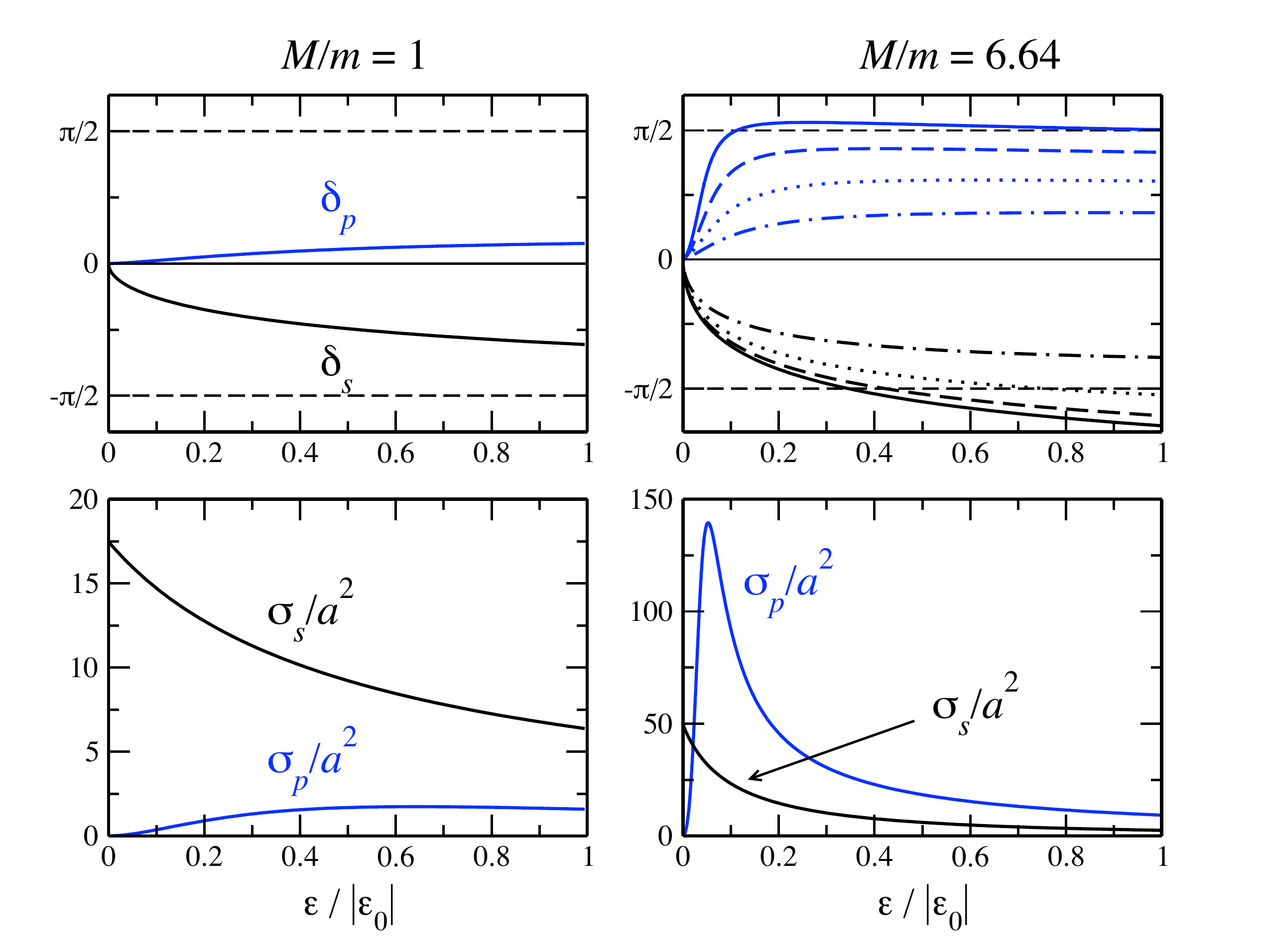}
\vspace{-4mm}
\caption{(color online). The scattering phase shifts and partial cross-sections as functions of the atom-dimer collision energy normalized to the dimer binding energy for the mass-balanced case (left) and for the case $M/m=6.64$ (right). In the top-right graph we also present the dependence on the width of the atomic resonance. Dashed, dotted, and dash-dotted lines stand for $a/R^*=16$, 4, and 1, respectively.}
\label{fig:phases}
\end{figure}

The singular and regular terms in Eq.~(\ref{PsiLimit}) should match the two-body scattering problem and, hence, are related by the Bethe-Peierls boundary condition $\Psi \propto (1/|{\bf r - R}_1|-1/\tilde a(E_c))$, where the energy dependent scattering length defined as $\tilde a(E_c)^{-1}=a^{-1}+2\mu R^* E_c$ is evaluated at the collision energy of the two atoms \cite{Petrov_v2_2004}. Here $a$ is the scattering length (at zero collision energy) and the length parameter $R^*$ is inversely proportional to the width of the Feshbach resonance. The collision energy, $E_c$, can be obtained by separating the relative motion of the colliding pair and subtracting the energies of all other degrees of freedom from the total energy of the system. Thus, $E_c$ can be written as a differential operator acting on $f$:
\begin{equation}\label{Ec}
\hat E_c f=\left[E+\frac{\nabla_{{\bf R}_1}^2}{2(M+m)}-\frac{(M\Omega^2+m\omega^2) Z_1^2}{2}-\hat H_2\right] f.\nonumber
\end{equation}
Finally, the integro-differential equation for $f$ reads
\begin{equation}\label{ResEq}
[-\hat K +a^{-1}+2\mu R^* \hat E_c] f({\bf R}_1,{\bf R}_2)=0.
\end{equation}
Equation (\ref{ResEq}) can be regarded as a two-body Schr\"odinger equation describing the motion of a dimer and an atom, $f({\bf R}_1,{\bf R}_2)$ being the corresponding wavefunction. 

Let us now consider the atom-dimer scattering in the uniform space, $\Omega=\omega=0$. In this case the center-of-mass motion separates, $f({\bf R}_1,{\bf R}_2)= f({\bf R}_1-{\bf R}_2)$, and it is convenient to work in the momentum representation, $f({\bf p})=\int f({\bf R})\exp(i{\bf pR}){\rm d}^3R$. Then $\hat E_c = E_c (p)= E-p^2/2\mu_3$ with $\mu_3=M(M+m)/(2M+m)$, and 
\begin{equation}\label{Kmomentum}
\hat K f({\bf p})\!=\!\sqrt{-2\mu E_c(p)}f({\bf p})+\!\int\!\!
\frac{f({\bf q})/(2\pi^2)\; {\rm d}^3q}{p^2\!+q^2\!+(2\mu/m){\bf p}{\bf q}-2\mu E}.\nonumber
\end{equation}
The calculation of the atom-dimer scattering amplitude from Eq.~(\ref{ResEq}) then follows the standard method developed for the ordinary Schr\"odinger equation in momentum space \cite{LLQ}. In Fig.~\ref{fig:phases} we present the resulting $s$- and $p$-wave phase shifts (top) and the corresponding partial scattering cross-sections (bottom) as functions of the atom-dimer collision energy $\varepsilon$ normalized to the dimer binding energy $|\varepsilon_0|$ defined by $\sqrt{2\mu |\varepsilon_0|}-a^{-1}+2\mu R^*|\varepsilon_0|=0$. We consider the atom-dimer scattering below the dimer breakup threshold, so that the total energy $E=-|\varepsilon_0|+\varepsilon<0$. On the left we present the results for the mass-balanced case and we see that the $s$-wave contribution to the scattering is always dominant. On the right, in the case $M/m=6.64$, solid lines show the phase shifts and scattering cross-sections in the limit of a wide resonance, or very small detunings ($a\gg R^*$). One can clearly see that the $p$-wave contribution dominates everywhere except for very small $\varepsilon$. The $p$-wave phase shift crosses the unitarity line $\delta_p=\pi/2$ at $\varepsilon\approx 0.1|\varepsilon_0|$ and stays very close to it at larger collision energies. 

On the top-right in Fig.~\ref{fig:phases} we present the scattering phase shifts in the case of a Feshbach resonance of finite width. We see that the narrower the resonance or the larger the detuning, the weaker the atom-dimer interaction (attractive $p$-wave as well as repulsive $s$-wave). This is consistent with the fact that the atoms forming the dimer spend more time in the closed channel state, interaction of which with the other atom is not resonant.

At this point we can conclude that for sufficiently small detunings $R^*/a$ a mixture of K atoms and K-Li dimers should behave quite differently compared to the case of an atom-dimer mixture in which dimers are composed of equal-mass atoms. The resonant character of the $p$-wave atom-dimer interaction can be demonstrated, for example, by colliding with each other two cold clouds of atoms and molecules and by measuring the angular distribution of scattered atoms or molecules \cite{LongPaper}.

We see that the atom-dimer scattering properties are sensitive to the detuning $R^*/a$. However, for the Li-K mass ratio, increasing this parameter can only weaken the atom-dimer interaction. Aside from the scaling with $|\varepsilon_0|$, is it possible to shift the position of the $p$-wave atom-dimer resonance down and eventually turn it into a trimer state by changing parameters of the system? According to Ref.~\cite{Kartavtsev2007} this can be achieved by increasing the mass ratio to the value $M/m\approx 8.2$. Another approach, not involving changes in atom masses, is based on the following facts. In the 2D case the K-K-Li system with zero-range interactions has one bound trimer state with unit angular momentum \cite{PricoupenkoPedri2008} and in the case of mixed dimensions, when K is two-dimensional and Li is three-dimensional, the 3-body system exhibits the Efimov effect; i.e., it supports an infinite number of trimer states \cite{Nishida1,Nishida2}. Therefore, one can assume that a gradual increase of an external quasi-2D confinement does shift the position of the atom-dimer resonance and turns it into a trimer state.

\begin{figure}[bt]
\includegraphics[width=0.8\hsize]{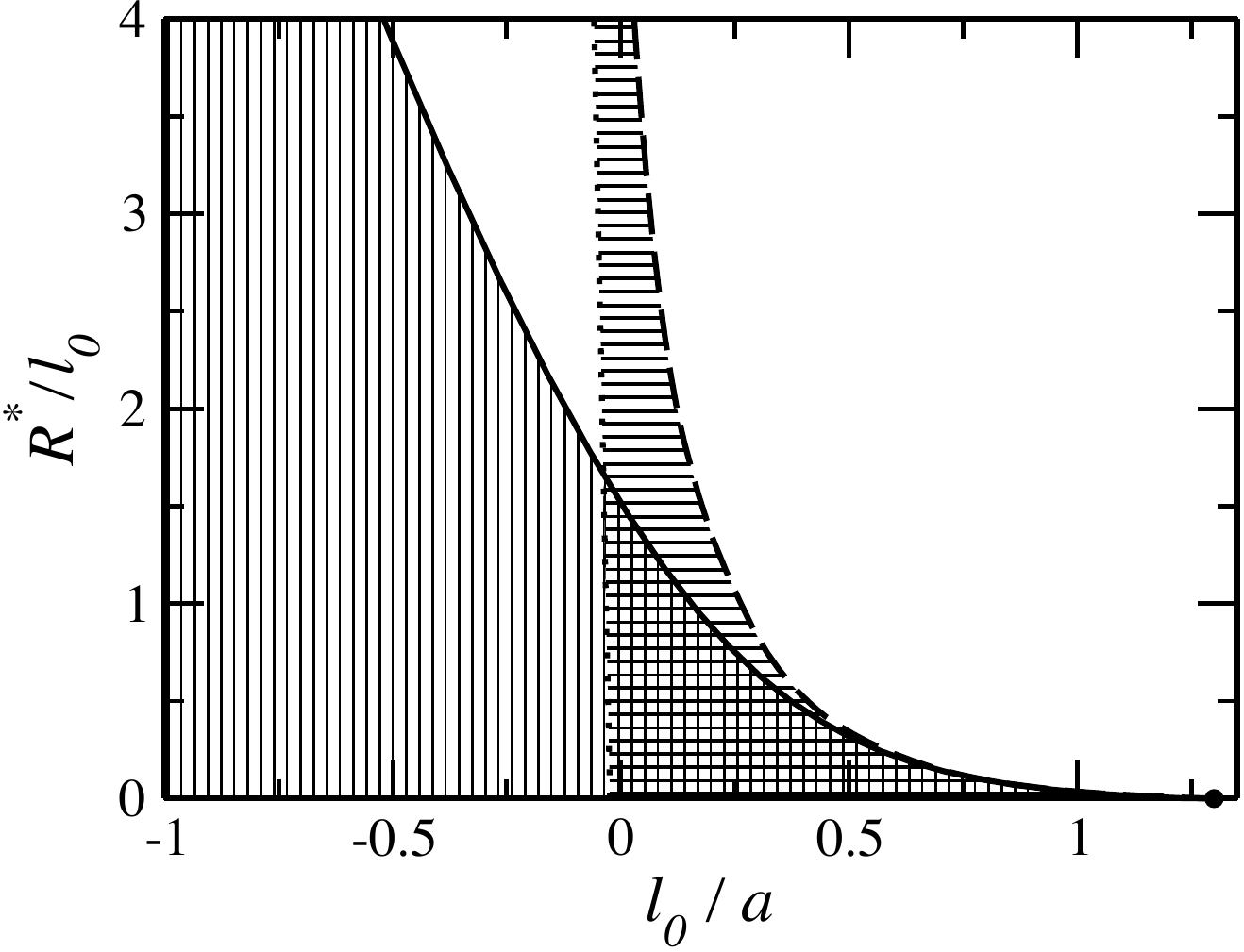}
\vspace{-2mm}
\caption{Phase diagram of the K-K-Li system in an external quasi-2D confinement. The solid (dashed) line corresponds to the trimer threshold in the case $\Omega=\omega$ ($\omega=0$). Three atoms form a bound state in vertically (horizontally) shaded areas. The dotted line represents the dimer formation threshold in the case $\omega=0$ (see text).}
\vspace{-5mm}
\label{fig:Thresholds}
\end{figure}

To answer the question of how strong should be the confinement we return to Eq.~(\ref{ResEq}) and solve it in the inhomogeneous case. Note that all the operators in Eq.~(\ref{ResEq}) conserve the planar center-of-mass momentum as well as the planar angular momentum $\ell$ and, therefore, $f$ essentially depends only on three coordinates: $f({\bf R}_1, {\bf R}_2)=f(Z_1,Z_2,|{\mbox{\boldmath$\rho$}}_1-{\mbox{\boldmath$\rho$}}_2|)\exp(i \ell \phi)$, where $Z_i$ and $\mbox{\boldmath$\rho$}_i$ are the axial and radial components of ${\bf R}_i$, and $\phi$ is the angle of ${\mbox{\boldmath$\rho$}}_1-{\mbox{\boldmath$\rho$}}_2$. Besides, in the case $\Omega=\omega$ the center-of-mass motion in the axial direction separates and the configuration space of Eq.~(\ref{ResEq}) becomes two-dimensional. 

Figure~\ref{fig:Thresholds} shows the trimer formation thresholds for $\omega=\Omega$ (solid line) and $\omega=0$ (dashed line), and we introduce the heavy atom oscillator length, $l_0=1/\sqrt{M\Omega}$. In both cases the critical confinement length for $R^*=0$ is given by $l_0/a\approx 1.3$ and then rapidly decreases with $R^*$. The closeness of the two curves for $a\lesssim 2l_0$ is explained by the fact that the dimer size, $\sim a$, is smaller than the light atom oscillator length, $l_0\sqrt{M\Omega/m\omega}$, and the dimer and trimer states are fairly insensitive to the light atom confinement. Accordingly, this part of the diagram should be quite universal for $0\leq\omega\lesssim \Omega$. 

On the negative side of the resonance the two cases differ significantly. For finite $\omega$, there always exists a dimer state whose size exponentially increases with the detuning $|l_0/a|$ ($a<0$). The problem then becomes essentially 2D (cf. \cite{PricoupenkoPedri2008}) and for sufficiently small $R^*$ there is one trimer state with unit angular momentum (vertically shaded area). In the case $\omega=0$ there exists a dimer threshold (dotted line) where the dimer size diverges and the resulting mixed-dimensional 2D-3D system exhibits the Efimov physics \cite{Nishida1,Nishida2}. We find, however, that the corresponding region in the vicinity of this line is 7 orders of magnitude narrower than the horizontally shaded area, which indicates a single trimer state.

Let us now discuss realistic values of $a$, $R^*$, and $l_0$, achievable in ongoing experiments. Hereafter we make $\hbar$ explicit. For the resonance at $B=114.47(5)$G in the $\left\vert 1/2,+1/2\right\rangle _{\mathrm{Li}}-\left\vert 9/2,+9/2\right\rangle_{\mathrm{K}}$ mixture, believed to be one of the widest of the $^{6}$Li-$^{40}$K system, the width is $\Delta B=1.5(5)$G \cite{Tiecke2009}. With the background scattering length $a_{\mathrm {bg}}=3$nm and $\mu_{\mathrm {rel}}=1.57\,\mu_{B}$ for the difference in magnetic moments of the closed and open channels we have \cite{Petrov_v2_2004} $R^*=\hbar^{2}/(2\mu a_{\mathrm {bg}}\mu_{\mathrm {rel}}\Delta B)\simeq100$nm. The same value of $l_0$ is achieved for the confinement frequency $\Omega\approx 2\pi\times 25$kHz. Then the trimer formation threshold is reached for $a\approx 400$nm. Note that $R_e\approx 2.2$nm and the zero-range approach is well justified.

It is well established \cite{p-wave} that atomic systems close to a $p$-wave resonance suffer from very strong losses due to the relaxation to deep molecular states. Because of the centrifugal barrier the size of a $p$-wave bound state is of the order of $R_e$, independent of the detuning. Therefore, $p$-wave molecules easily relax to deeper bound states of approximately the same size when colliding with another atom or molecule. The two-body relaxation rate constant is then of the order of $\alpha_{\mathrm {rel}}\propto \hbar R_e/m$. Translating this picture to our $p$-wave atom-dimer resonance we observe that the range of the atom-dimer potential, given mostly by the exchange of the light atom, is one or two orders of magnitude larger than $R_e$. However, the relaxation rate constant $\alpha_{rel}$ does not scale accordingly as there is only a single trimer state and there are no deeper states of the same size to which it can relax. Thus, the main loss channel in the gas should be the relaxation to molecular states of the size $R_e$. This mechanism is responsible for the decay of a single trimer, the rate of which we will now discuss. It is also important for the atom-dimer mixture close to the $p$-wave resonance as atom triples spend a lot of time in this bound or quasi-bound trimer state.

Consider a K-K-Li trimer in the case $R^*=0$ and $a\sim l_0$. For hyperradii $R_e\ll \xi \ll a$ its wavefunction reads $\Psi \approx \xi^{\nu_p-1}/a^{\nu_p+2}\Phi(\hat\Omega)$, where $\Phi(\hat\Omega)$ is a normalized function of hyperangles and $\nu_p\approx 0.198$ corresponds to the heavy-heavy-light three-fermion problem with $M/m=6.64$ and $p$-wave symmetry \cite{Petrov2003,Lifetime}. The decay rate equals $1/\tau=P(R_e)\nu_{\mathrm {mic}}$, where $P(R_e)\approx (R_e/a)^{2\nu_p+4}$ is the probability to find all three atoms in the microscopic region $\xi\lesssim R_e$, and $\nu_{\mathrm {mic}}\sim \hbar/mR_e^2$ represents the order of magnitude estimate of the relaxation rate once the atoms are confined to this region (cf. \cite{Nishida2}). In the case of finite $R^*$ it follows from the coupled-channel analysis that the open-channel wavefunction $\Psi({\bf R}_1,{\bf R}_2,{\bf r})\approx f({\bf R}_1,{\bf R}_2)/4\pi |{\bf r - R}_1|$ is always accompanied by the closed-channel amplitude $\sqrt{R^*/4\pi}f({\bf R}_1,{\bf R}_2)$, and estimates of the normalization integral calculated with these functions show that the latter is dominant at $\xi\sim |{\bf R}_1-{\bf R}_2|\ll R^*$. This means that at $\xi \ll R^*$ the three-atom problem reduces to the problem of a non-interacting atom and a closed-channel (bare) molecule. Accordingly, in the case $R_e\ll R^*\ll a$ the trimer decay rate is a product of three factors, $1/\tau=P(R^*)\nu_{\mathrm {mic}}(R_e/R^*)^5$, where the last factor gives the probability for a bare molecule and an atom with unit total angular momentum to approach each other from distances $\sim R^*$ to distances $\sim R_e$. Similar analysis can be performed for other values of $a$, $R^*$, and $l_0$, and substituting realistic numbers for Li-K mixtures one obtains that the trimer decay rate is 3 to 5 orders of magnitude lower than the dimer binding frequency and can lead to trimer lifetimes of order 1s. In contrast to the Efimov case, the binding of these purely long-range trimers comes from distances of the order of the dimer size, whereas at short distances the three-atom dynamics is dominated by the centrifugal barrier. We see that the $p$-wave atom-dimer resonance does not compromise the lifetime of the system.

Independent of the proximity of the $p$-wave atom-dimer resonance the lifetime of the mixture is likely to be limited by the relaxation in $s$-wave atom-dimer and dimer-dimer collisions. A detailed discussion of these processes in the case $R^*=0$ can be found in Ref.~\cite{Lifetime} and can be generalized to the case of finite $R^*$ in the same way as outlined above. In particular, in the case $R_e \ll R^* \ll a$ the 3D $s$-wave atom-dimer relaxation rate constant equals $\alpha_{\mathrm {rel}}\approx (\hbar R_e/m) (R^*/a)^{5.04}$, and we get a similar power law dependence for the dimer-dimer channel with a slightly lower exponent. We see that for small detunings from the Feshbach resonance; i.e., for $R^*\ll a$, these rates are strongly suppressed. Thus, also in this respect the mixture of K atoms and K-Li dimers holds promise as a unique long-lived system in which resonant $p$-wave interactions can be observed. 

We thank V. Gurarie for fruitful discussions. This work was supported by the IFRAF Institute, by ANR (grant 06-NANO-014), by the EuroQUAM-FerMix program, by the Dutch Foundation FOM, and by the Russian Foundation for Fundamental Research.


\vspace{-2mm}

\begin{thebibliography}{99}
\vspace{-2mm}

\bibitem{Review} See for review, {\it Ultracold Fermi Gases}, Proceedings of the International School of Physics ``Enrico Fermi'', edited by M.~Inguscio, W.~Ketterle, and C.~Salomon (IOS Press, Amsterdam, 2008).

\bibitem{Zwierlein2006} M.~W.~Zwierlein {\it et al.}, Science {\bf 311}, 492 (2006).

\bibitem{Partridge2006} G.~B.~Partridge {\it et al.}, Science {\bf 311}, 503 (2006).
 
\bibitem{Zwierlein2005} M.~W.~Zwierlein {\it et al.}, Nature (London) {\bf 435}, 1047 (2005).

\bibitem{Taglieber2007} M.~Taglieber {\it et al.}, Phys. Rev. Lett. {\bf 100}, 010401 (2008).

\bibitem{Wille2008} E.~Wille {\it et al.}, Phys. Rev. Lett. {\bf 100}, 053201 (2008).

\bibitem{Tiecke2009} T.~G.~Tiecke {\it et al.}, e-print arXiv:0908.2071.

\bibitem{Voigt2009} A.-C.~Voigt {\it et al.}, Phys. Rev. Lett. {\bf 102}, 020405 (2009).

\bibitem{Schreck2009} F.~Schreck, talk at ICTP, Trieste, 2009.

\bibitem{Jin2008} K.-K.~Ni {\it et al.}, Science {\bf 322}, 231 (2008).

\bibitem{Aymar2005} M.~Aymar and O.~Dulieu, J. Chem. Phys. {\bf 122}, 204302 (2005).

\bibitem{Petrov2007} D.~S.~Petrov {\it et al.}, Phys. Rev. Lett. {\bf 99}, 130407 (2007).

\bibitem{Efimov1973} V. N. Efimov, Nucl.~Phys.~A {\bf 210}, 157 (1973).

\bibitem{Lifetime} D.~S.~Petrov, C.~Salomon, and G.~V.~Shlyapnikov, J.~Phys.~B: At.~Mol.~Opt.~Phys.~{\bf 38} S645 (2005).

\bibitem{Kartavtsev2007} O.~I.~Kartavtsev and A.~V.~Malykh, J. Phys. B: At. Mol. Opt. Phys. {\bf 40}, 1429 (2007).

\bibitem{Petrov2003} D.~S.~Petrov, Phys.~Rev.~A {\bf 67}, 010703(R) (2003).

\bibitem{Petrov_v2_2004} D.~S.~Petrov, Phys.~Rev.~Lett.~{\bf 93}, 143201 (2004).

\bibitem{Mora2004} C.~Mora {\it et al.}, Phys. Rev. Lett. {\bf 93}, 170403 (2004).

\bibitem{LLQ} L.~D.~Landau and E.~M.~Lifshitz, {\it Quantum Mechanics}, (Butterworth-Heinemann, Oxford, 1999).

\bibitem{LongPaper} J.~Levinsen {\it et al.}, to be published.

\bibitem{PricoupenkoPedri2008} L.~Pricoupenko and P.~Pedri, e-print arXiv:0812.3718.

\bibitem{Nishida1} Y.~Nishida and S.~Tan, Phys. Rev. Lett. {\bf 101}, 170401 (2008).

\bibitem{Nishida2} Y.~Nishida and S.~Tan, Phys. Rev. A {\bf 79}, 060701(R) (2009).

\bibitem{p-wave} C.~A.~Regal {\it et al.}, Phys. Rev. Lett. {\bf 90}, 053201 (2003); J.~Levinsen, N.~R.~Cooper, and V.~Gurarie, {\it ibid.} {\bf 99}, 210402 (2007), Phys. Rev. A {\bf 78}, 063616 (2008); M.~Jona-Lasinio, L.~Pricoupenko, and Y.~Castin, Phys. Rev. A {\bf 77}, 043611 (2008). 

\end{thebibliography}
\end{document}